\documentclass[10pt,journal]{IEEEtranTCOM}
\usepackage[english]{babel}
\usepackage[usenames]{color}
\usepackage[cp1250]{inputenc}
\usepackage{amsfonts}
\usepackage{amsthm}
\usepackage{graphicx}
\usepackage{epsfig}
\usepackage{mathrsfs}
\usepackage{amsmath}
\usepackage{algorithm}
\usepackage{algorithmic}
\usepackage{hyperref}

\theoremstyle{plain}

\begin{document}
\newcommand{\bea}{\begin{eqnarray}}
\newcommand{\eea}{\end{eqnarray}}
\newcommand{\be}{\begin{equation}}
\newcommand{\ee}{\end{equation}}
\newcommand{\beas}{\begin{eqnarray*}}
\newcommand{\eeas}{\end{eqnarray*}}
\newcommand{\bs}{\backslash}
\newcommand{\bc}{\begin{center}}
\newcommand{\ec}{\end{center}}
\def\SC {\mathscr{C}}

\title{Encoding of probability distributions\\for Asymmetric Numeral Systems}
\author{\IEEEauthorblockN{Jarek Duda}\\
\IEEEauthorblockA{Jagiellonian University,
Golebia 24, 31-007 Krakow, Poland,
Email: \emph{dudajar@gmail.com}}}
\maketitle

\begin{abstract}
Many data compressors regularly encode probability distributions for entropy coding - requiring minimal description length type of optimizations. Canonical prefix/Huffman coding usually just writes lengths of bit sequences, this way approximating probabilities with powers-of-2. Operating on more accurate probabilities usually allows for better compression ratios, and is possible e.g. using arithmetic coding and Asymmetric Numeral Systems family. Especially the multiplication-free tabled variant of the latter (tANS) builds automaton often replacing Huffman coding due to better compression at similar computational cost - e.g. in popular Facebook Zstandard and Apple LZFSE compressors. There is discussed encoding of probability distributions for such applications, especially using Pyramid Vector Quantizer(PVQ)-based approach with deformation, bucket approximation, prefix trees, improving accuracy with additional bits, also tuned symbol spread for tANS.
\end{abstract}
\section{Introduction}
Entropy/source coding is at heart of many of data compressors, transforming sequences of symbols from estimated statistical models, into finally stored or transmitted sequence of bits. Many especially general purpose compressors, like widely used Facebook Zstandard\footnote{https://en.wikipedia.org/wiki/Zstandard} and Apple LZFSE\footnote{https://en.wikipedia.org/wiki/LZFSE}, regularly store probability distribution for the next frame e.g. for byte-wise $D=256$ size alphabet for e.g. $L=2048$ state tANS automaton: tabled Asymmetric Numeral Systems (\cite{ANS1,ANS2,ANS3}).

\begin{figure}[t!]
    \centering
        \includegraphics[width=8.5cm]{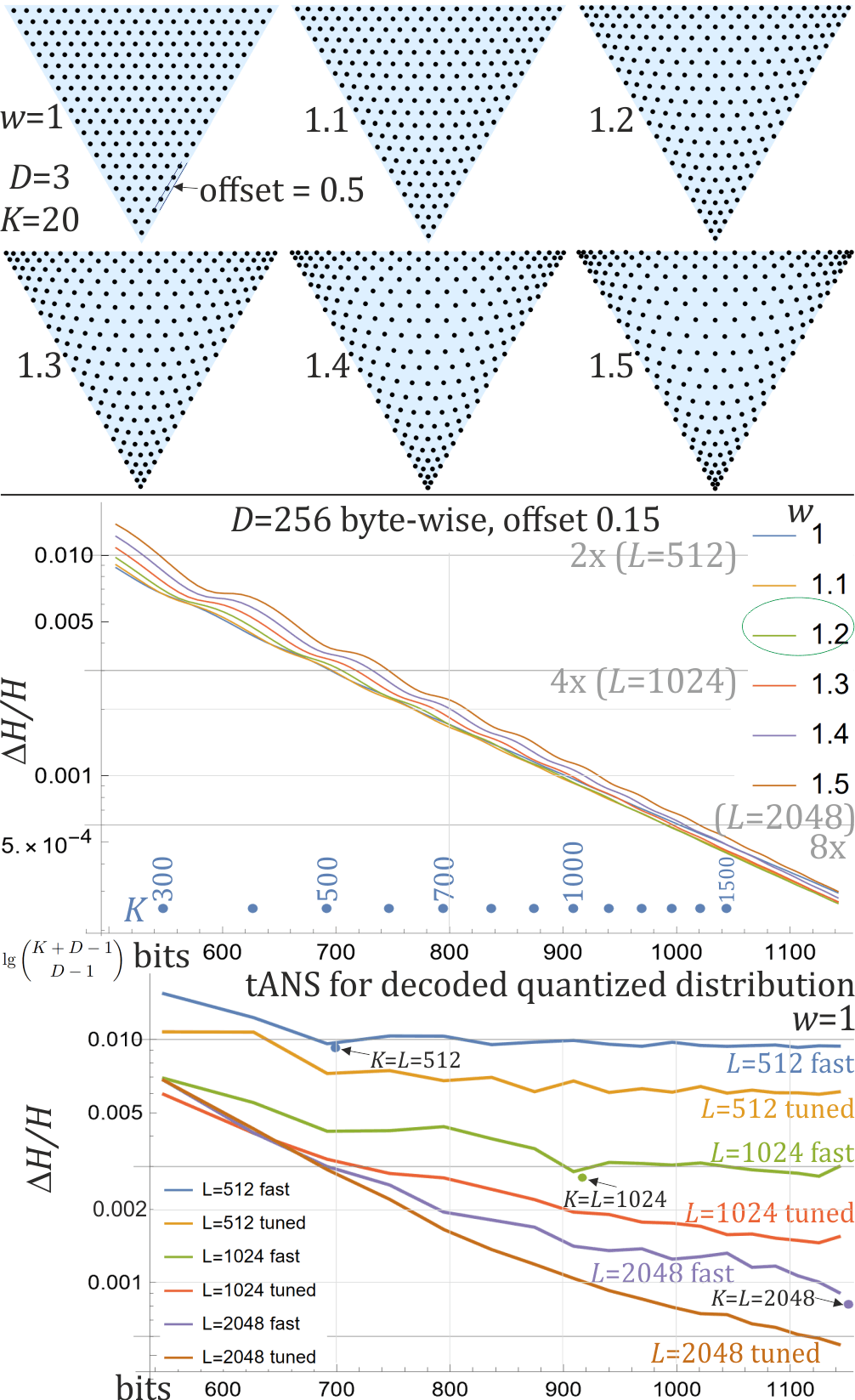}
        \caption{Top: probability quantization codebooks for $D=3$ variables and $K=20$ sum and various used powers $w$ to increase accuracy for low probabilities. Offset shifts from the boundary to prevent zero probabilities.
        Middle: $\Delta H/H$ increase of size due to discussed quantization (vertical, averaged over 1000 randomly chosen distributions) for various costs of encoding of probability distribution  (horizontal, or used sum $K$ in blue) for encoding of byte-wise $D=256$ size alphabet as in Zstandard or LZFSE. Plots of various colors correspond to powers $w$ used for deformation as in below diagrams. The three horizontal lines are for $\Delta H/H\in \{0.01, 0.003, 0.0006\}$ approximately corresponding to tANS with $2, 4, 8$ times more states than alphabet size $D$ from \cite{ANS2}. Bottom: $\Delta H/H$ evaluation of tANS automaton built for such encoded probabilities (for $w=1$ power) using fast spread as in Zstandard, and slightly better tuned spread discussed here. The three marked dots correspond to directly storing quantized probabilities for tANS fast spread. E.g. 800 bit header allows to work $\Delta H/H\approx 0.0015$. }
        \label{pq}
\end{figure}

This article focuses on  optimization of encoding of such probability distribution for this kind of situations, Fig. \ref{pq} summarizes its results. For historical Huffman coding there are usually used canonical prefix codes~\cite{canon} for this purpose - just store lengths of bit sequences. However, prefix codes can only process complete bits, approximating probabilities with powers-of-2, usually leading to suboptimal compression ratios. ANS family has allowed to include the fractional bits at similar computational cost, but complicating the problem of storing such more accurate probability distributions - required especially for inexpensive compressors.

Assuming the source is from $(p_s)_{s=1..D}$ i.i.d. probability distribution, the number of (large) length $N$ symbol sequences (using $n!\approx (n/e)^n$ Stirling approximation, $\lg\equiv \log_2$) is:
$${N\choose Np_1,Np_2,\ldots,Np_D}=\frac{N!}{(Np_1)!\ldots (Np_D)!}\approx 2^{N H(p)}$$
\be \textrm{for Shannon entropy:}\quad H(p)=-\sum_s p_s \lg(p_s) \ee
So to choose one of such sequence, we need asymptotically $\approx H(p)$ bits/symbol, what can imagined as weighted average: \emph{symbol of probability $p$ carries $\lg(1/p)$ bits of information} - entropy coders should indeed approximately use on average.

However, if entropy coder uses $(q_s)_{s=1..D}$ probability distribution instead, e.g. being powers-of-2 for prefix/Huffman coding, it uses $\lg(1/q_s)$ bits for symbol $s$ instead, leading to Kullback-Leibler (KL) $\Delta H$ more bits/symbol:
\be \Delta H_{p,q}=\sum_s p_s \lg\left(\frac{p_s}{q_s} \right)\approx \frac{1}{\ln 4} \sum_s \frac{(p_s-q_s)^2}{p_s}\label{KL}\ee
where approximation is from Taylor expansion to 2nd order.

Therefore, we should represent probability distribution optimizing MSE (mean-squared error), but weighted with $1/p$: the lower probability of symbols, the more accurate representation should be used.

Additionally, there is some cost $c(q)$ bits of storing this probability distribution - we want to optimize here. Finally, for length $N$ sequence (frame of data compressor), we should search for minimum description length~\cite{MDL} type of compromise:
\be\textrm{minimize penalty:}\qquad \arg\min_{q}\quad c(q) + N\ \Delta H_{p,q} \ee
The longer frame $N$, the more accurately we should represent $p$. Additionally, further entropy coding like tANS has own choices (quantization, symbol spread) we should have in mind in this optimization, preferably also including computational cost of necessary operations in the considerations.

The main discussed approach was inspired by Pyramid Vector Quantization (PVQ)~\cite{PVQ} and possibility of its deformation~\cite{PVQ1} - this time for denser quantization of low probabilities to optimize (\ref{KL}). Its basic version can be also found in \cite{reznik}, here with multiple optimizations for entropy coding application, also tANS tuned spread~\cite{toolbox}.

Later version of this article has added approaches with initial approximation: using buckets or prefix trees, then adding bits to improve accuracy where it is the most beneficial (could be also applied for PVQ approach).
\section{Encoding of probability distribution}
This main section discusses quantization and encoding of probability distribution based of PVQ as in Fig. \ref{pq}. There is a short subsection about handling zero probability symbols. Fig. \ref{q1} briefly presents alternative approach, discussed in the next Section.

\subsection{Zero probability symbols}
Entropy coder should use on average $\lg(1/q)$ bits for symbol of assumed probability $q$, what becomes infinity for $q=0$ - we need to ensure to prevent this kind of situations.

From the other side, using some minimal probability $q_{min}$ for $k$ unused symbols, means the remaining symbols can use only total $1-kq_{min}$, requiring e.g. to rescale them $q_s \to q_s (1-kq_{min})$, what means:
\be -\lg(1-kq_{min})\approx k q_{min} \quad \textrm{bits/symbol penalty} \label{pen} \ee
Leaving such unused symbols we would need to pay this additional cost multiplied by sequence length $N$.

To prevent this cost, we could mark these symbols as unused and encode probabilities of the used ones, but pointing such $D-k$ out of $D$ symbols is also a cost - beside computational, e.g. $\lg {D \choose k}\approx D h(k/D)$ bits for $h(p)=-p \lg(p)-(1-p)\lg(1-p)$.

For simplicity we assume further that all symbols have nonzero probability, defining minimal represented probability with offset $o$. Alternatively, we could remove this offset if ensuring that zero probability is assigned only to unused symbols.

\subsection{Probabilistic Pyramid Vector Quantizer (PPVQ)}
Assume we have probability distribution on $D$ symbols with all nonzero probabilities - from interior of simplex:
$$S_D =\{p\in \mathbb{R}^D: \sum_s p_s=1, \forall_s\, p_s > 0\}$$

The PVQ~\cite{PVQ} philosophy assumes approximation with fractions of fixed denominator $K$:
$$p_s \approx Q_s/K \quad\textrm{for}\quad Q_s \in \mathbb{N},\ \sum_s Q_s = K$$

Originally PVQ also encodes sign, but here we focus on natural numbers. The number of such divisions of $K$ possibilities into $D$ coordinates $n(D,K)$ is the number of putting $D-1$ dividers in $K+D-1$ positions - the numbers of values between these dividers are our $Q_s$:
$$n(D,K) := {K+D-1 \choose D-1}\approx 2^{(K+D-1)h((D-1)/(K+D-1))}$$
for $h(p)=-p\lg(p)-(1-p)\lg(1-p)$ Shannon entropy.

To choose one of them, the standard approach is enumerative coding: transform e.g. recurrence $(k=Q_1+\ldots+Q_s)$
$$n(s,k)=\sum_{Q_s=0,\ldots,k} n(s-1,k-Q_s),\qquad n(1,k)=1$$
into encoding: first shifting by positions to be used by lower $Q_s$ possibilities, then recurrently encoding further:
$$\textrm{code}(Q_s Q_{s-1}..Q_1)=\textrm{code}(Q_{s-1}..Q_1)+\sum_{i=0}^{Q_s-1} n(s-1,k-i)$$
However, it would require working with very large numbers, e.g. in hundreds of bits for standard $D=256$ byte-wise distributions. Additionally, this encoding assumes uniform probability distribution of all possibilities, while in practice it might be very nonuniform, optimized for various data types. Hence, the author suggests to use entropy coding e.g. rANS instead of enumerative coding - prepare coding tables for all possible $(s,k)$ cases, using probabilities in basic case from division of numbers of possibilities:
\be \textrm{Pr}(Q_s =i|s,k)=\frac{n(s-1,k-i)}{n(s,k)} \ee
This way also easily allows for deformation of probability distribution on $S_D$ simplex, e.g. low probabilities might be more likely - what can be included for example by taking some power of theses probabilities and normalize to sum to 1. Such optimization will require further work e.g. based on real data, depending on data type (e.g. 3 different for separate in Zstandard: offset, match length, and literal length).

\subsection{Deformation to improve KL}
The discussed quantization uses uniform density in the entire simplex $S_D$, however, the $\Delta H$ penalty is $\propto \sum_s (p_s-q_s)^2/p_s$, suggesting to use denser quantization in low probability regions. While such optimization is a difficult problem, which is planned for further work e.g. analogously to adaptive quantization in \cite{aq}, let us now discuss simple inexpensive approach as in \cite{PVQ1}, presented in Fig. \ref{pq}.

Specifically, it just uses power $w$ of quantized probabilities - search suggests to use $w\approx 1.2$, additionally we need some offset to prevent zero probabilities - which generally can vary between symbols (for specific data types), but for now there was considered constant $o_s=o$, optimized by $o\approx 0.15$:
\be q_s = \frac{(Q_s)^w + o_s}{\sum_{s'} (Q_{s'})^w + o_{s'}} \ee
for quantization we need to apply $1/w$ power $p_s \to (p_s)^{1/w}$ before searching for the fractions:
\be \frac{Q_s}{K} \approx \frac{(p_s)^{1/w}}{\sum_{s'} (p_{s'})^{1/w}} \ee
In theory we should also subtract the offset here, but getting improvement this way will require further work.

\subsection{Finding probability quantization}

Such quantization: search for $Q\in \mathbb{N}^D$ for which $Q/K$ approximates chosen vector, is again a difficult question - testing all the possibilities would have exponential cost. Fortunately just multiplying by the denominator $K$ and rounding, we already get approximate $Q$ - there only remains to ensure $\sum_s Q_s = K$ by modifying a few coordinates, what can be made minimizing KL $\propto (p_s-q_s)^2/p_s$ penalty, also ensuring not to get below zero.

Such quantizer optimizing KL penalty can be found in \cite{toolbox} for C++, here there was used one in Wolfram Mathematica, also with example of used evaluation:

\begin{footnotesize}
\begin{verbatim}
guard = 10.^10;        (*to prevent negative values*)
quant[pr_]:=(pk =K*pr; Qs =Round[pk]+0.;kr=Total[Qs];
   While[K != kr, df = Sign[K - kr];
    mod = Sign[Qs + df];   (*modifiable coordinates*)
    penalty=((Qs+df-pk)^2-(Qs-pk)^2)/pk+guard(1-mod);
    best=Ordering[penalty,Min[Abs[kr-K],Total[mod]]];
    Do[Qs[[i]] += df, {i, best}]; kr=Total[Qs]];Qs );

(* example evaluation: *)
d = 256; n = 1000; K = 1000; w = 1.2; offset = 0.15;
ds = RandomReal[{0,1},{n, d}];ds=Map[#/Total[#]&,ds];
Qd = Map[quant, Map[#/Total[#] &, ds^(1/w)]];
qt = Map[#/Total[#] &, Qd^w + offset];
dH = Map[Total,ds*Log[2.,qt]]/
 Map[Total,ds*Log[2.,ds]]-1;
Mean[dH]
\end{verbatim}
\end{footnotesize}

\section{Approximate then improve CDF encoding} \label{s2}

\begin{figure}[t!]
    \centering
        \includegraphics[width=8.5cm]{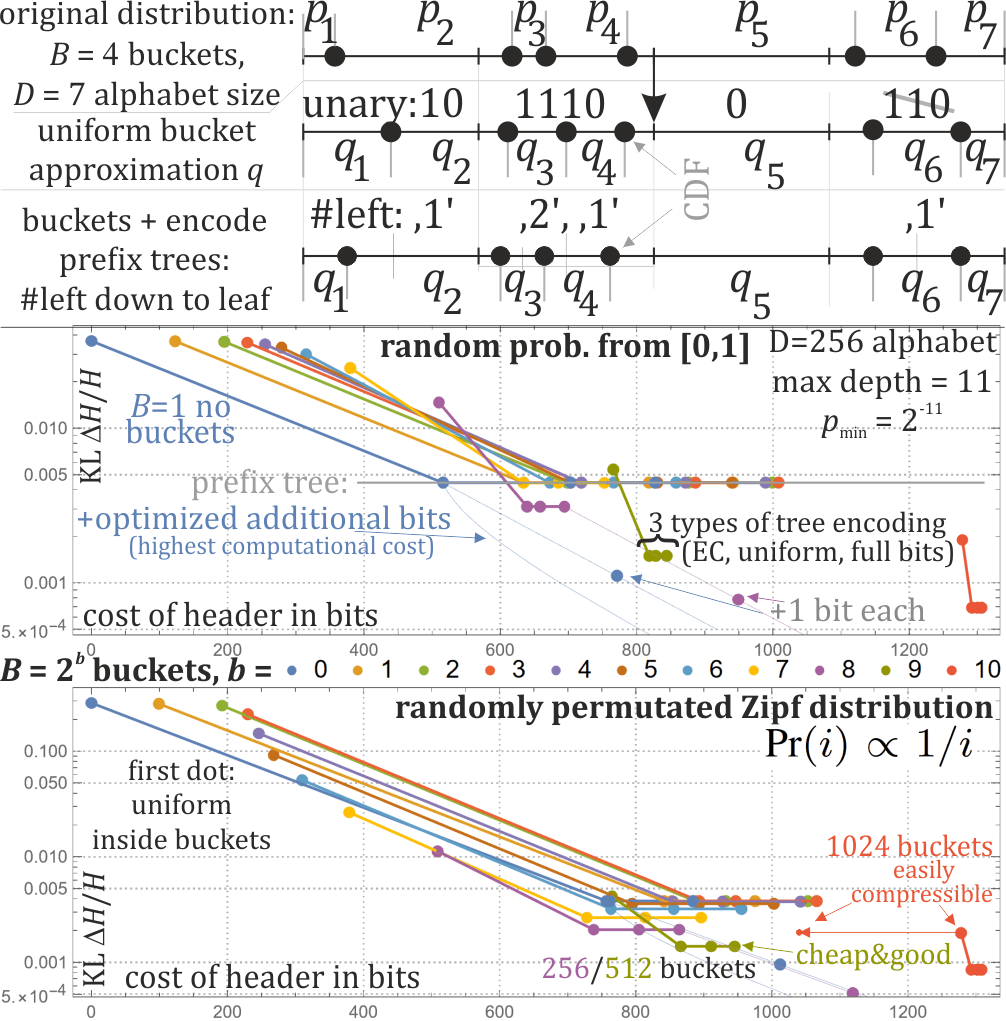}
        \caption{Section \ref{s2} approach. \textbf{Top}: CDF $c_i=\sum_{j<i} p_j$ for $i=1..D-1$ (dots) are encoded with bucket trick: the range is split into $B=2^b$ buckets (skipped for $B=1$), for all but the last buckets there was encoded number of values inside with unary coding. Then multiple elements in buckets are just distributed uniformly, or there is further encoded prefix tree inside: written how many elements go left and so on recursively until reaching leaf.
        \textbf{Middle/bottom} - its tests for $D=256$ alphabet size: KL only $\Delta H/H$ vs header cost, for two distributors: up - probabilities chosen randomly from [0,1] then normalized, down: randomly permutated Zipf distribution $\textrm{Pr}(i)\propto 1/i$. First probabilities below $p_{min}=2^{-11}$ were increased to this value (for $L=2^{11}=2048$ states, with tuning should be chosen $p_{min}=2^{-12}$). Then there was calculated CDF and its $D-1$ values are encoded with discussed approach for $B=2^b$ buckets and $b=0,\ldots,10$. Each has four dots: upper corresponds to just spreading uniformly inside, 3 lower dots to further encoding prefix tree in 3 ways (entropy coder, uniform distribution, full bits). Then additional bits can be used: "+1 bit each" costs $D-1=255$ bits and $\approx 4$ times improves $\Delta H/H$. Thin line also shows adding bits in optimized positions (neighboring low probability symbols), what allows for slight improvement. Full prefix tree gives $\Delta H/H\approx 0.004$, obtained also for low $B$.  }
        \label{app}
\end{figure}
In this section we will discuss alternative approach as in Fig. \ref{app} originally (buckets only) proposed by the author in 2014\footnote{\url{https://encode.su/threads/1883}}. For CDF (cumulative distribution function):
\be C_0=0,\ C_D=1, \quad\textrm{ for }i=1..D-1:\quad C_i := \sum_{j<i} p_i \ee
first encode its approximation $c_i\approx C_i$ ($c_0=0, c_D=1, q_i=c_i-c_{i-1}$) with discussed further bucket (very low computational cost) and/or prefix tree (higher cost and performance) approach. Then eventually a sequence of additional bits - for all CDF values, or better neighboring low probable symbols.

Discussed next bucket and tree approximations, which can be used separately or combined, can bring decoder both approximated values $c_i$ (positive increasing), if wanting to use additional bits also error estimates $e_i$: ensuring that real $C_i\in [c_i - e_i, c_i +e_i)$.

As discussed further, often it is convenient to have certainty that $p\geq p_{min}$, for example $p_{min}=1/L$ for $L$ state tANS, or $p_{min}=0.5/L$ with tuning. If not satisfied, it can be enforced as initial approximation: increase to $p_{min}$ all values below, and rescale the remaining to sum up to 1.

\subsection{Initial approximation: buckets with unary coding}
Let us start with the original approach, inspired by nice bucket approximation from James Dow Allen\footnote{\url{http://fabpedigree.com/james/appixm.htm}}.

Imagine we want to write $n$ elements in $\{1,\ldots,L\}$ range, originally e.g. hash values, here CDF. Directly it would require $n \lg(L)$ bits. However, here we do not need their order, allowing to save $\lg(n!) \approx n \lg(n/e)$ bits (Stirling approximation), finally requiring $\approx n(\lg(L/n)+1.443)$ bits.

We could do it with entropy coder, but it is relatively costly. The bucket approximation trick is splitting the $\{1,\ldots,L\}$ range into $B=n$ bucket (further we will use general $B$, preferably $L,B$ power-of-2) as subranges. Then for each bucket encode with unary coding the number of elements in this buckets (what finally requires $B$ of '0' and $n$ of '1'), then encode the suffixes for each element ($\approx n\lg(L/n)$ bits), finally requiring $\approx n(\lg(L/n)+2)$ bits - what is $\approx 0.557n$ bits worse than optimum.

We can reduce this $\approx 0.557n$ penalty to $\approx 0.142n$ by replacing unary coding with trit: '0','1' means that there is 0,1 element in this bucket. In contrast, '2' means there is 2 or more elements, using their order to encode the exact number, e.g. sorting them and exchanging the last two - this way decoder knows to close the bucket when the read element is in reversed order.\\

Returning to CDF encoding of $D-1$ values in $\{1,\ldots,L\}$, we can use unary coding as in Fig. \ref{app}, for numbers of buckets conveniently being a power-of-2 $B=2^b$ as $L$ usually also is - the larger, the better (initial) approximation, but also higher bit cost. Knowing the alphabet size e.g. $D=256$, we can skip encoding of the number of elements in the last bucket.

Now for buckets with a single appearance we can choose $c_i$ as its center. For $k>1$ elements in the bucket, we can e.g. split the bucket into $k$ identical subranges, with $c_i$ as their centers - referred as \textbf{uniform bucket approximation}, the lowest cost dots in evaluation in Fig. \ref{app}. Better accuracy at higher header cost is further encoding distribution inside $k>1$ buckets with prefix tree below, maybe also further additional bits improving precision.

Ensuring $p\geq p_{min}$ assumption, we get convenient bound for $k\leq 1/(B p_{min})$. E.g. for $B=1024$ buckets and $L=2048$ states, using $k\leq 2$ we directly get probabilities as required, for $k\leq 4$ we can use tANS tuning: for $k=3,4$ shift right the singletons. We can also use discussed further tuned spread, which can approximate well any probabilities. 

The $B=1024$ buckets case has only $\approx 1/4$ nonempty buckets here, allowing for reduction with simple data compression e.g. just entropy coder for $\textrm{Pr}(0)=3/4, \textrm{Pr}(1)=1/4$ would reduce it $1278\to \approx 1037$ bits.
\subsection{Prefix tree alone ($B=1$) or with buckets ($B>1$)}
\begin{figure}[t!]
    \centering
        \includegraphics[width=8.5cm]{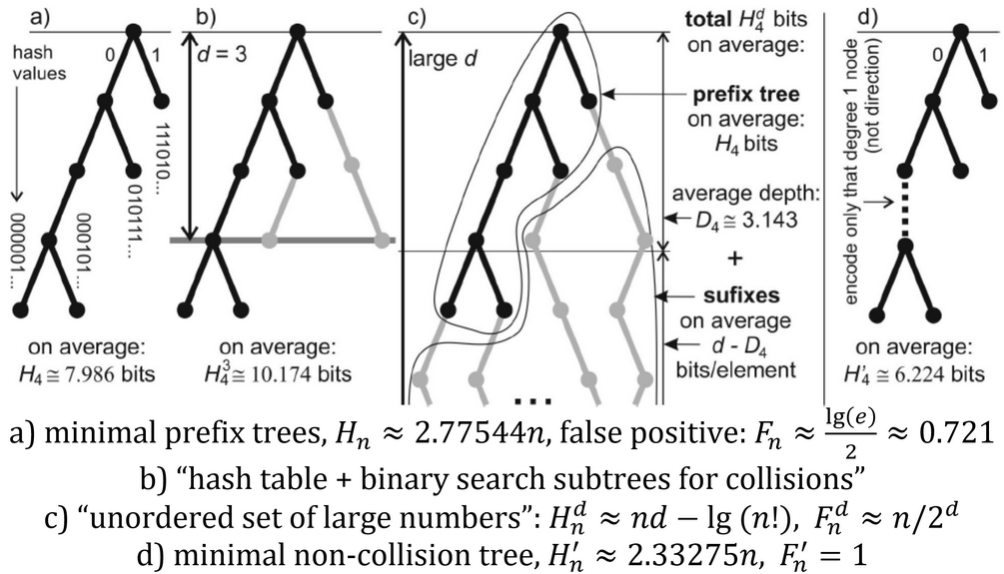}
        \caption{Asymptotic entropy behavior for storing minimal prefix trees required to distinguish $n$ random bit sequences form \cite{hash}. For storing approximated CDF as their binary expansion, we are interested in the most-left a) case, requiring asymptotically $\approx 2.77544n$ bits (using entropy coder), additionally being restricted by maximal depth recognized by entropy coder ($\lg(L)$, +1 with tuning, automatically bounded by $\lg(1/p_{min})$ if ensuring $p\geq p_{min}$) like in b), maybe plus additional bits similarly to c), but this time based on KL evaluation.}
        \label{hash}
\end{figure}

Another considered initial approximation (also for above buckets with $k>1$ elements), is encoding minimal prefix tree of bit sequences as binary expansions of $(c_i)$ CDF values -  encoding minimal numbers of bits sufficient to distinguish from other values, like in Fig. \ref{hash}. We can do straightforward ($B=1$) for CDF in $[0,1]$, or combined with above buckets: in size $1/B$ subranges, e.g. skipping first $b$ bits in expansion for $b=2^b$.

Decoder usually knows the number of elements for tree e.g. $D=256$, hence $D-1$ CDFs (or $k>1$ numbers of elements in bucket). A natural way to encode a tree is first to encode for the root: how many elements are in its left subtree. In our case, how many of $(c_i)_{i=1..D-1}$ start with digit '0' in binary expansion. Then do the same for the two subtrees (asking for the second digit), and so on in-order or pre-order encoding the tree, until the number of elements in the current subtree drops to 1 (leaf). This approximation looks convenient as giving more accurate representations in low probability regions.

The difficulty is encoding "how many goes left" for each internal node. For $l$ elements in current  subtree, $k\in\{0,\ldots,l\}$ of them in its left subtree, assuming uniform distribution we get: $\textrm{Pr}(k)={l \choose k}/2^l$. Doing so, as discussed in \cite{hash} and presented in Fig. \ref{hash}, asymptotically we would need $\approx 2.77D $ bits to encode the tree. However, it would require entropy coder e.g. tANS and preparing all these probability distributions. We could also approximate distribution using the fact that ${l \choose k}/2^l$ is $\approx$ Gaussian distribution centered in $l/2$ and of variance $l/4$ - allowing e.g. to write the high bits using some fixed Gaussian distribution, and the low bits directly.

If instead of entropy coder, to reduce cost we would just write $k\in\{0,\ldots,l\}$ using $\lg(l)$ bits, what assumes uniform $k$ distribution (can be done with numeral systems or e.g. tANS with approximate probabilities), the cost grows from $\approx 2.8D$ to $\approx 3.2D$ bits. Even worse if using  $\lceil\lg(l)\rceil$ bits to work on complete bits - the cost grows to approx $3.7D$ bits. Finally the 3 lower dots in \ref{app} evaluation show the 3 possibilities of tree encoding: differing by $\sim 100$ bits for tree-only ($B=1$), this difference shrinks with the number of buckets - resolving similar value localization problem in slightly different way.

Finally if $C_i$ was encoded with $b_i$ bits (depth of its leaf, $+b$ with buckets), the decoder has its approximation $c_i$ as center of corresponding range, with error estimation as half of its width:
\be c_i=(\lfloor C_i 2^{b_i} \rfloor+1/2)\, 2^{-b_i} \qquad\qquad  e_i =2^{-b_i -1} \label{tr}\ee

For extremely low probable symbols, such tree could have arbitrary depth, what means additional cost not useful for entropy coder, e.g. for tANS with $\approx 0.72/L$ smallest represented probability. It suggests to use mentioned initial approximation enforcing $p\geq p_{min}$ for $p_{min}=1/L$ or $0.5/L$ with some tANS tuning (e.g. placing $p < 1/L$ symbols as singletons at the end of symbol spread). 

In theory we could alternatively stop expanding tree if reaching some depth, and finally spread uniformly if there are still multiple CDFs in this maximal depth. However, it still would need further probability quantization (avoided if $p\geq 1/L$), and cannot use discussed further additional bits.

Enforcing $p\geq p_{min}$ by initial approximation, we automatically bound the maximal depth of the tree to $-\lg(p_{min})$. We have certainty that $C_i\in [c_i - e_i, c_i +e_i)$ separate for each $C_i$, what allows to additionally read more bits to increase accuracy.

\subsection{Additional bits to improve approximation}
If from above we have separate subranges $C_i\in [c_i - e_i, c_i +e_i)$, we could read further bits to improve accuracy - e.g. reading 1 more bit all ($D-1$ bits), we would reduce $e_i$ twice, reducing $\Delta H/H$ approximately four times.

As thin lines in Fig. \ref{app}, this asymptotic behavior can be slightly improved if first reading bits defining low probabilities (having larger influence on KL).

Specifically, imagine decoder has received some approximation of CDF: $c_i\approx C_i$ with error control: certainty that $C_i \in [c_i-e_i,c_i+e_i)$. They are encoded for $i=1,\ldots,D-1$, additionally taking $c_0=0, c_D=1, e_0=e_D=0$. Such approximation could come from the bucket and/or prefix tree approximation, but also from discussed PVQ ($e_i$ as half of width), or maybe different approaches - e.g. from previous data frame, now with separate additional bits optimized for the current frame.

Assuming uniform distribution in such $C_i \in [c_i-e_i,c_i+e_i)$ range, mean squared error of $p_i \approx q_i=c_i-c{i-1}$ probability can be estimated as
$$ \frac{1}{4e_{i-1}e_i}\int_{-e_{i-1}}^{e_{i-1}} dx \int_{p_i-e_i}^{d+e_i} dy\ (y-x-p_i)^2 =\frac{1}{3} (e_{i-1}^2 +e_i^2)$$
Combining with (\ref{KL}) and using $q_i=c_i-c_{i-1}$ approximated distribution, decoder can estimate KL as
\be \Delta H\approx \frac{1}{3 \ln 4} \sum_{i=1}^D \frac{e_{i-1}^2 +e_i^2}{q_i} \ee
Reading additional bit for $c_i$ would reduce $e_i\to e_i/2$, reducing redundancy $\Delta H \to^\approx \Delta H - g_i$ by estimated gain $g_i$:
\be g_i = \frac{1}{4 \ln 4} \left(\frac{1}{q_i} + \frac{1}{q_{i+1}}\right)\, e_i^2 \label{gs} \ee
While the $q_i$ approximation would improve after reading such bit and this update could be applied, a natural approach is calculating it ones (rough approximation is sufficient), then just reduce it $g_i \to g_i/4$ after reading the corresponding bit.

For length $N$ sequence (data frame), ignoring entropy coder inaccuracy, such gain would mean $\approx N g_i$ bits saved at cost of 1 additional bit describing probability distribution. Hence this bit would be worth storing/reading if $g_i > 1/N$, what can be used evaluating from decoder perspective. Assuming such bit reduces $g_i$ four times, for given $i$ we should use additional $\lceil \log_4 (Ng_i)\rceil$ bits.

In practice, to reduce cost, we can e.g. choose some low probability bound, and for symbols of lower probabilities read one additional bit for the two neighboring CDF values.
\subsection{Summary of buckets, trees, additional bits}
In this section there were discussed various approaches with complex dependencies. Here is a brief summary based on in Fig. \ref{app} evaluation:
\begin{itemize}
  \item Usually the lowest header cost is for just encoding prefix tree, then optimized additional bits. However, it is computationally costly,
  \item For low computational cost, the 256, 512, 1024 buckets cases are the most promising. Especially in the 1024 case such header can by easily reduced by further compression e.g. just entropy coder,
  \item It is beneficial to initially enforce/approximate $p\geq 1/L$ (or $0.5/L$ with tuning),
  \item The behavior is dependent on probability distribution family - it might be worth to have implemented e.g. 2 ways and choose one based on the probability distribution.
\end{itemize}

\section{Tuned symbol spread for tANS}
\begin{figure}[b!]
    \centering
        \includegraphics[width=8.5cm]{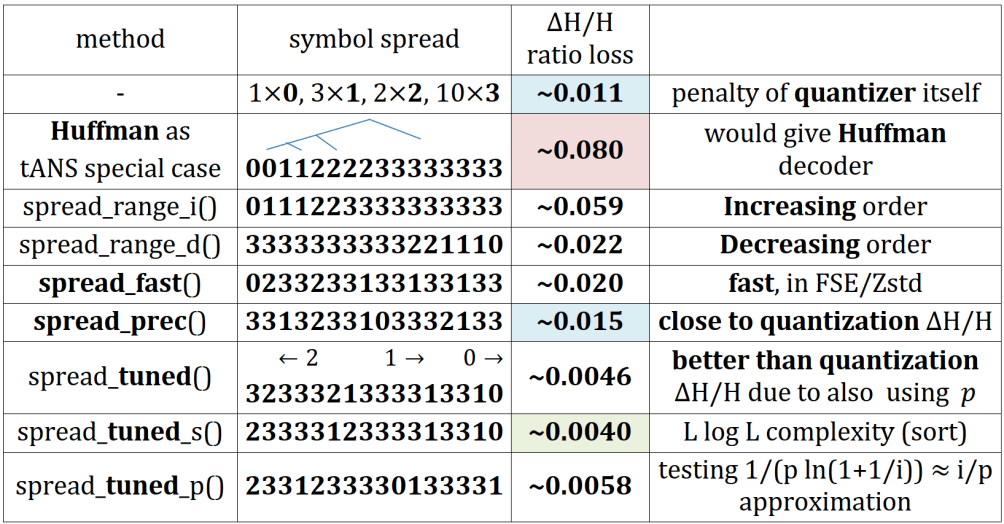}
        \caption{Example from~\cite{toolbox} of tANS $\Delta H/H$ compression ratio penalty (from Shannon entropy $H$) for various symbol spreads, $L=16$ states, $D=4$ size alphabet with $p=(0.04,0.16,0.16,0.64)$ probabilities quantized to $\textbf{L}=(1,3,2,10)$ approximating probabilities with $\textbf{L}/16=(0.0625,0.1875,0.125,0.625)$. This quantization itself has penalty (first row). Spreading symbols in power-of-2 ranges we would get Huffman decoder for the drawn tree. The \textit{spread\_fast()} is as in Zstandard and LZFSE. A bit closer to quantization error \textit{spread\_prec()} can be found in \cite{ANS2}. Here we discuss \textit{spread\_tuned()} which, beside quantization, uses also the original probabilities $p$ - shifting symbol appearances toward left (increasing probability e.g. of '2' here as quantized $0.125$ is smaller than preferred $0.16$) or right (decreasing probability e.g. of '0' and '1' here), this way "tuning" symbol spread to be even better than quantization penalty.}
        \label{spreads}
\end{figure}
While the discussed quantization might be used for arithmetic~\cite{AC}  or rANS entropy coders, the most essential application seems tANS, where again we need quantization e.g. using $L=2048$ denominator for $D=256$ alphabet in Zstandard and LZFSE. There is another compression ratio penalty from this quantization, heuristics suggest  $\Delta H \sim (D/L)^2$ general behavior~\cite{ANS2}, approximately $\Delta H \approx 0.01$ for $L=2D$, $\Delta H \approx 0.003$ for $L=4D$, $\Delta H \approx 0.0006$ for $L=8D$ - these three cases are marked in Fig. \ref{pq}.

It is tempting to just use the same quantization for both: $K=L$, also presented in Fig. \ref{pq}, what basically requires $\approx 8 D h(1/8)\approx 4.35 D$ bits in $K=L=8D$ case ($\approx 139$ bytes for $D=256$) - the discussed deformation rather makes no sense in this case, zero probability symbols can be included. Cost of such header can be reduced e.g. if including nonuniform distribution on $S_D$ simplex, or using some adaptivity.

We can reduce this cost by using $K<L$ and deformation, Fig. \ref{pq} suggests that essential reduction should have nearly negligible effect on compression ratios. However, it would become more costly from computational perspective.

Also, if using deformation, we need second quantization for the entropy coder. Let us now briefly present tuned symbol spread introduced by the author in implementation~\cite{toolbox}, here first time in article.

The tANS variant, beside quantization: $p_s \approx L_s/L$ approximation, also needs to choose symbol spread: $\overline{s}:\mathcal{A}\to \{L,\ldots,2L-1\}$ with $L_S$ appearances of symbol $s$, where $\mathcal{A}$ is the alphabet of size $|\mathcal{A}|=D$ here.

Intuitively, this symbol spread should be nearly uniform, but finding the best one seems to require exponential cost, hence in practice there are used approximations - Fig. \ref{spreads} contains some examples, e.g. shifting modulo by constant and putting appearances of symbols - fast method used in Finite State Entropy\footnote{https://github.com/Cyan4973/FiniteStateEntropy} implementation of tANS used e.g. in Zstandard and LZFSE. Here is source used for Fig. \ref{pq}:

\begin{footnotesize}
\begin{verbatim}
fastspread := (step = L/2+L/8+3; spr=Table[0,L];ps=1;
   Do[Do[ps = Mod[ps + step, L]; spr[[ps + 1]] = i
    ,{j, Ls[[i]]}], {i, d}]; spr);
\end{verbatim}
\end{footnotesize}
A bit smaller penalty close to $p_s \approx L_s/L$ quantization is precise spread~\cite{ANS2} focused on indeed being uniform.
\subsection{Tuned symbol spread - using also probabilities}
Given symbol spread, assuming i.i.d $(p_s)$ source, we can calculate stationary probability distribution of states (further (\ref{ss})) $x\in I=\{L,\ldots,2L-1\}$ symbols, which is usually approximately $\rho_x\equiv\textrm{Pr}(x)\approx \lg(e)/x$.

\begin{figure}[t!]
    \centering
        \includegraphics[width=8.5cm]{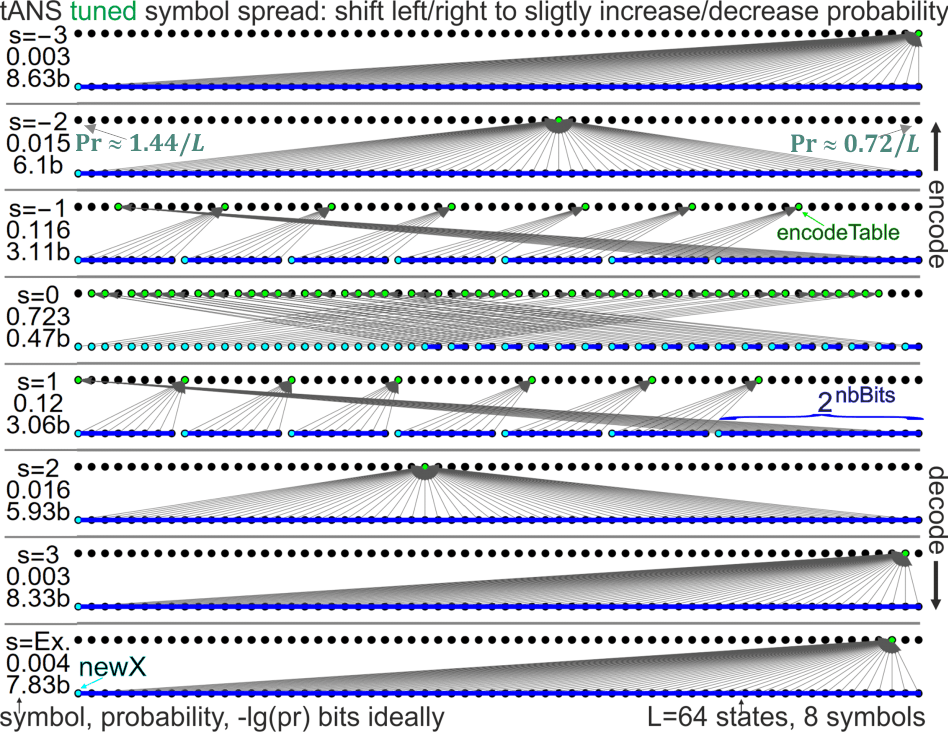}
        \caption{Example of tuned tANS automaton for $D=8$ size alphabet and $L=64$ states. The $\textrm{Pr}(x)\approx \lg(e)/x$ state probability distribution for $x\in \{L,\ldots,2L-1\}$ allows e.g. for singletons ($L_s=1$) to represent probability from $\approx 0.72/L$ (most-right) to $\approx 1.44/L$ (most-left). As in KL inaccuracy of the lowest probable symbols is the most important, a basic tuning used in FSE marks the very low probable symbols and places them in most-right position, done above for $s\in \{-3,3,Ex\}$ symbols. Used here full tuning additionally places the remaining symbols in nearly uniform way, and shifts them left/right to slightly increase/decreare from represented $L_s/L$ probability ($s$ has $L_s$ appearances).}
        \label{tn}
\end{figure}

\begin{figure}[t!]
    \centering
        \includegraphics[width=8.5cm]{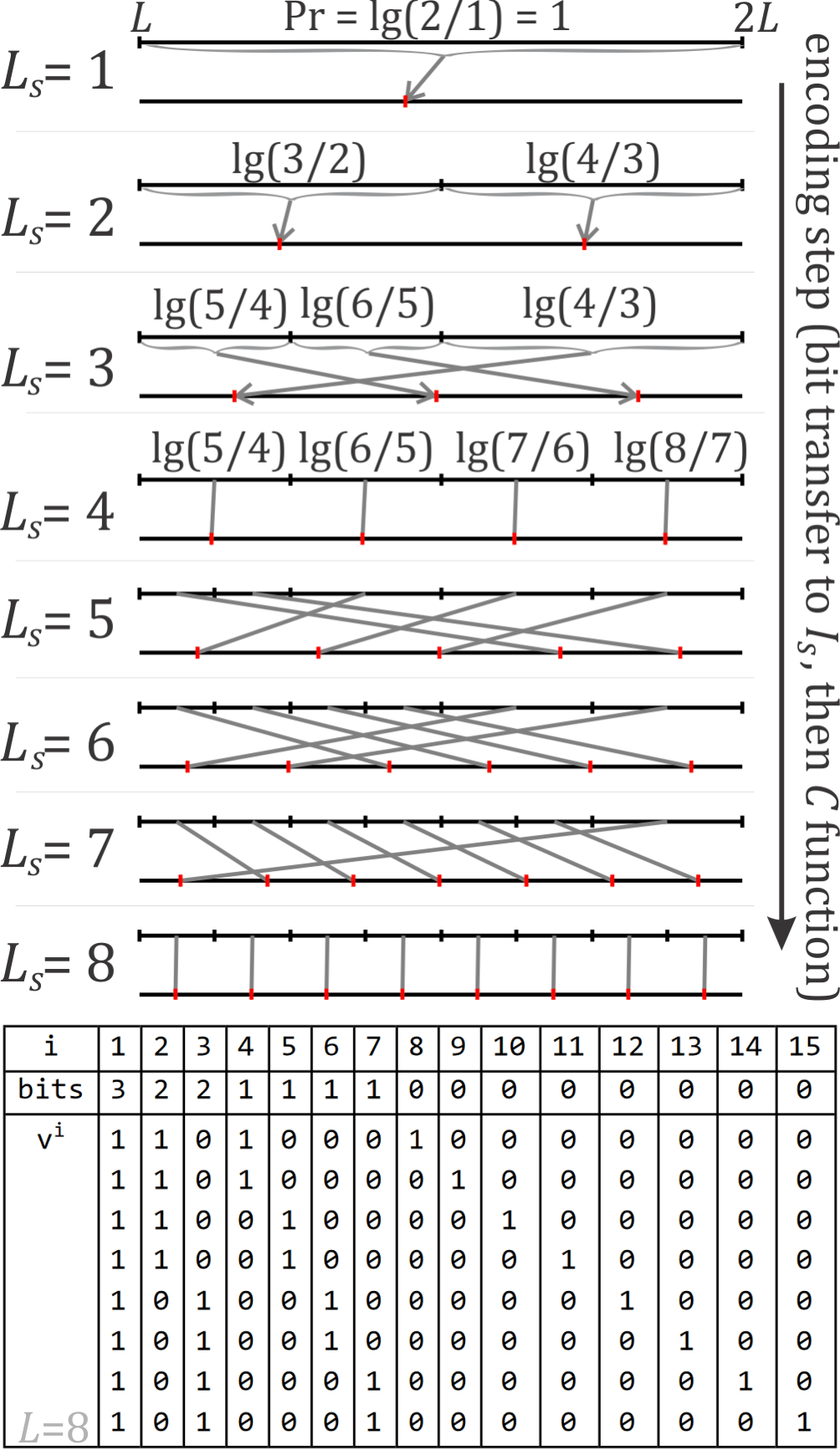}
        \caption{Top: Diagram for tANS tuned symbol spread: behavior of stream encoding step for $L_s=1\ldots 8$ symbol appearances, approximating probabilities with $\approx L_s/L$ for $L=2^l$ number of states. There are first written bits determining position inside one of $L_s$ ranges, then there is transition to one of $L_s$ points marked red. Using $\textrm{Pr}(x)\approx \lg(e)/x$ probability distribution of states, we get the written approximate probability distributions of ranges: $\textrm{Pr}(i)\approx \lg((i+1)/i)$ for $i=L_s,\ldots,2L_s-1$. Multiplying these probabilities by assumed probabilities of symbols $p_s$, and comparing with $\textrm{Pr}(x)\approx \lg(e)/x$ approximation, we get the preferred marked red positions: to $x \approx 1/(p_s \ln(1+1/i))$ - tuned symbol spreads try to use. Bottom: these used ranges for $L=8$ in form of vectors and number of bits in renormalization - which e.g. allows to quickly build the stochastic matrix.  }
        \label{tuned}
\end{figure}

Encoding symbol $s$ from state $x$ requires first to get (e.g. send to bitstream) some number of the least significant bits: $\mod(x,2)$, each time reducing state $x\to \lfloor x/2\rfloor$, until $x$ gets to $\{L_s,\ldots, 2L_s -1 \}$ range. Then such reduced $x$ is transformed into corresponding appearance in $\{x\in I: \overline{s}(x)=s\}$ - in Fig \ref{tuned} the upper ranges correspond to the same reduced $x$, finally transformed into one of $L_s$ marked red positions.

This way we split $I$ into $L_s$ power-of-2 size subranges. Approximated $\textrm{Pr}(x)\approx \lg(e)/x$ probability distribution of states, combined with $\sum_{i=1}^k 1/i \approx \ln(k) +\gamma$ harmonic number approximation ($\gamma\approx 0.577$), allows to find simple approximation for probability distribution of such ranges: $$\textrm{Pr}(i\textrm{-th subrange})\approx \lg((i+1)/i)=\lg(1+1/i)$$

Multiplying such (approximate) subrange probability by probability of symbol $p_s$, we get probability of the final state, which should agree with $\textrm{Pr}(x)\approx \lg(e)/x$. Equalizing these two approximate probabilities we get:

\be x\approx \frac{1}{p_s} \frac{1}{\ln(1+1/i)}\quad\textrm{for } i=L_s,\ldots 2L_s-1 \ee
preferred positions for $s$ in symbol spread $\overline{s}$.

Hence we can gathered all (symbol, preferred position) pairs with $L_s$ appearances of symbol $s$, sort all these pairs by the preferred position, and fill $I=\{L,\ldots,2L-1\}$ range accordingly to this order - this approach is referred as \textit{spread\_tuned\_s()} in \cite{toolbox} and Fig. \ref{spreads}.\\

Here is the used source - prepares pairs of $(x,symbol)$, sort by the first coordinate, then take the second coordinate:
\begin{footnotesize}
\begin{verbatim}
(* Ls, pd - quantized, assumed distribution *)
tunedspread := Transpose[Sort[Flatten[Map[Transpose[
 {1/Log[1.+1/Range[#[[2]], 2#[[2]]-1]] /#[[3]],
 Table[#[[1]], #[[2]]]}] &,
 Transpose[{Range[d], Ls, pd}]], 1]]][[2]];
\end{verbatim}
\end{footnotesize}

However, above sorting would have $O(L \lg(L))$ complexity, we can reduce it to linear e.g. by rounding preferred positions to natural numbers and putting them into one of such $L$ buckets, then go through these buckets and spread symbols accordingly - this is less expensive, but slightly worse \textit{spread\_tuned()}.

This table also contains approximation for logarithm, but in practice we should just put $1/\ln(1+1/i)$ for $i=1,\ldots,2\max_s(L_s)-1$ into a further used table.

\subsection{Calculation of tANS mean bits/symbol}
Assuming i.i.d. $(p_s)$ source, Fig. \ref{tuned} perspective allows to write optimization of symbol spread in form:
\be\min_{\Pi} \left\{\|\Pi B S\rho\|_1: \Pi S\rho=\rho,\,\rho>0,\,\sum_x \rho_x=1\right\}\ \frac{\textrm{bits}}{\textrm{symbol}} \label{ss}\ee
where $S$ is the initial stochastic matrix made of $p_s v^i$ rows for $i=L_s\ldots 2L_s-1$ (in order) and all symbols $s$. Symbol spread is defined by $\Pi$ - permutation matrix maintaining this $i=L_s\ldots 2L_s-1$ order for each symbol (cannot change order inside positions corresponding to one symbol). $B$ is diagonal matrix with numbers of bits (\verb"bits" in Fig. \ref{tuned}) in the same order as in $S$: $\lg(L)-\lfloor \lg(i)\rfloor$ bits for all these $i=L_s\ldots 2L_s-1$. The $\Pi S\rho=\rho$ condition (or equivalently $(\Pi S-I)\rho =0$)is to make $\rho$ the stationary probability distribution of used states.

The discussed tuned symbol spread can be seen as assuming $\rho^0_x =\lg(e)/x$ for $x=L,\ldots, 2L-1$ approximated stationary probability distribution of states, and sorting $S\rho^0$ coordinates $(\approx p_s \lg(1+1/i))$ in decreasing order to get permutation $\Pi$. We could iterate this process, what might lead to a bit better "iterated tuned symbol spread": find $\rho^1$ stationary distribution for tuned spread, then get new $\Pi$ by sorting $S\rho^1$, and use such symbol spread or perform further iterations.

Generally finding the optimal symbol spread seems to require exponential cost, but such iterative tuning, or maybe just searching permutations of neighboring symbols, could lead to some improvements - practical especially for optimizations of tANS automata for fixed distributions.\\

The above view was used here to calculate bits/symbol with below source - first prepare number of bits  \verb"nbt" and rows \verb"vi" to quickly build the stochastic matrix \verb"sm" of jumps over states of automaton (using real probabilities, not the encoded ones), obtaining permutation from symbol spread (\verb"is"). We can calculate its stationary probability distribution as kernel $((\Pi S-I)\rho =0)$, then $\|\Pi B S\rho\|_1$ gives mean used bits/symbol:

\begin{footnotesize}
\begin{verbatim}
(* prepare vi and nbt tables *)
prepL := (fr = 0; len = L;     (*[fr+1,fr+len] range*)
 nbt = Table[Log[2, L], {i, 2 L - 1}]; sub = 0;
 vi = Table[cur = Table[0., L];
  Do[cur[[i]] = 1., {i, fr + 1, fr + len}]; fr += len;
  nbt[[i]]-=sub; If[fr == L, fr = 0; len /= 2; sub++];
  cur, {i, 2 L - 1}];)
(*caluclate mean bits/symbol, rd - real distribution*)
entrspread:=(cL=Ls; is = Table[cL[[i]]++, {i,spread}];
  sm = rd[[spread]]*vi[[is]];  (* stochastic matrix *)
  rho = NullSpace[sm - IdentityMatrix[L]][[1]];
  rho /= Total[rho];            (* stationary distr.*)
  Total[(nbt[[is]]*sm).rho])    (* mean bits/symbol *)
\end{verbatim}
\end{footnotesize}

\section{Conclusions and further work}
There was presented PVQ-based practical quantization and encoding for probability distributions especially for tANS, also the best practical known to the author symbol spread for this variant.

This is initial version of article, further work is planned, for example:

\begin{itemize}
  \item In practice there is rather a nonuniform probability distribution on the simplex of distributions $S_D$ - various practical scenarios should be tested and optimized for, like separately for 3 different data types in Zstandard.
  \item Also we can use some adaptive approach e.g. encode difference from distribution in the previous frame e.g. using CDF (cumulative distribution function) evolution like: \verb"CDF[s] += (mixCDF[s] - CDF[s])>>rate", where \verb"mixCDF" vector can be calculated from the previous frame or partially encoded, \verb"rate" is fixed or encoded forgetting rate. The discussed approximations (low $K$ PVQ denominator, buckets, prefix tree) can be used as such \verb"mixCDF" for probability update.
  \item The used deformation by coordinate-wise powers is simple, but might allow for better optimizations e.g. analogous to adaptive quantization in \cite{aq}, what is planned for further investigation.
  \item Choice of offsets optimizations, handling of very low probabilities, might allow for further improvements.
  \item Combined optimization with further entropy coder like tANS might be worth considering, also fast and accurate optimization of symbol spread.
\end{itemize}

%\appendix
%\section{Adaptive quantization of probability distribution}
%Let us briefly sketch optimization of such deformation analogously to adaptive quantization from \cite{aq} - optimizing some function
%$$\textrm{vol}(S_D)=\int_0^1 \int_0^{1-p_D} \ldots dp_1\ldots dp_D$$
%$$\rho_{d,r}(p) = (r-p)^{d-1}$$
%$$\mathcal{D}_{dr}=\int_0^r \frac{\rho_{dr}(p)}{p\,q_{dr}(p)^2} dp$$

\bibliographystyle{IEEEtran}
\bibliography{cites1}
\end{document}